\documentclass[pdflatex,sn-nature]{sn-jnl}

\usepackage{graphicx}%
\usepackage{multirow}%
\usepackage{amsmath,amssymb,amsfonts}%
\usepackage{amsthm}%
\usepackage{mathrsfs}%
\usepackage[title]{appendix}%
\usepackage{xcolor}%
\usepackage{textcomp}%
\usepackage{manyfoot}%
\usepackage{booktabs}%
\usepackage{algorithm}%
\usepackage{algorithmicx}%
\usepackage{algpseudocode}%
\usepackage{listings}%
\usepackage{float}%
\usepackage{cleveref}%
\usepackage{enumerate}%

\begin{document}

\title[Article Title]{Judgement Citation Retrieval using Contextual Similarity}


\author*[1]{\fnm{Akshat Mohan} \sur{D}}\email{dasulaakshat@gmail.com}
\equalcont{These authors contributed equally to this work.}

\author*[2]{\fnm{Hrushitha} \sur{T}}\email{hrushithatigulla@gmail.com}
\equalcont{These authors contributed equally to this work.}

\author*[3]{\fnm{Preethika} \sur{B}}\email{bhukyapreethika23@gmail.com}
\equalcont{These authors contributed equally to this work.}

\affil*[1]{\orgdiv{Student in CSE (AIML)}, \orgname{CVR College of Engineering}, \orgaddress{\city{Ibrahimpatnam}, \postcode{501510}, \state{Telangana}}}

\affil*[2]{\orgdiv{Student in CSE (AIML)}, \orgname{CVR College of Engineering}, \orgaddress{\city{Ibrahimpatnam}, \postcode{501510}, \state{Telangana}}}

\affil*[3]{\orgdiv{Student in CSE (AIML)}, \orgname{CVR College of Engineering}, \orgaddress{\city{Ibrahimpatnam}, \postcode{501510}, \state{Telangana}}}


\abstract{Traditionally in the domain of legal research, the retrieval of pertinent citations from intricate case descriptions has demanded manual effort and keyword-based search applications that mandate expertise in understanding legal jargon. Legal case descriptions hold pivotal information for legal professionals and researchers, necessitating more efficient and automated approaches. We propose a methodology that combines natural language processing (NLP) and machine learning techniques to enhance the organization and utilization of legal case descriptions. This approach revolves around the creation of textual embeddings with the help of state-of-art embedding models. Our methodology addresses two primary objectives: unsupervised clustering and supervised citation retrieval, both designed to automate the citation extraction process. Although the proposed methodology can be used for any dataset, we employed the Supreme Court of The United States (SCOTUS) dataset, yielding remarkable results. Our methodology achieved an impressive accuracy rate of 90.9\%. By automating labor-intensive processes, we pave the way for a more efficient, time-saving, and accessible landscape in legal research, benefiting legal professionals, academics, and researchers.}




\maketitle

\section{Introduction}\label{sec1}

With the increasing prevalence of computers in law courts and public prosecutor offices, there is a massive and continuously growing volume of electronic text in the legal domain \cite{moens1997abstracting}. This vast amount of data, coupled with the inherent complexity of legal texts, makes it challenging to manage. As a result, there is an urgent need for advanced tools to process legal documents effectively, meeting the information needs of legal practitioners \cite{raghav2015text}. Therefore, with this huge load of legal data, it becomes difficult for a legal practitioner to extract citations and to quote them accurately making sure that they are directly relevant to their arguments to support their work. It is necessary to develop efficient search approaches that the users who can be a fresher practitioner or an experienced attorney without having to manually extract keywords from the current legal case and perform an exhaustive, time-consuming manual search.

This requires a thorough examination of each potential case to ensure that it is aligned with the legal problem at hand. They frequently struggle with the time-consuming task of scouring extensive databases to locate relevant cases. This process can be susceptible to oversight. To tackle these challenges and assist legal research this paper introduces an innovative application that utilizes text similarity algorithms. The fundamental purpose is to improve the efficiency and precision of legal research, allowing practitioners to acquire relevant material swiftly and precisely. \cite{chhatwal2018explainable}\cite{wei2018empirical}

In this application, legal opinions are analyzed and citations will be extracted by comparing them with a vast database of legal cases using the advanced text similarity algorithms. By ranking these citations based on similarity, the application can provide valuable recommendations for further analysis, it has the potential to significantly reduce the time and financial resources typically required for legal research. This enables legal professionals to allocate their efforts more strategically, focusing on higher-priority tasks rather than on identifying citations manually.

Additionally, the application fosters broader access to legal knowledge by making legal information more accessible. This will enable access to a wider array of participants, including those lacking extensive legal expertise, to actively engage in legal research and analysis. The versatility of the application is not limited to the legal profession but it may also be used in academic research, policy formulation, document preparation, and client consulting. 
Moreover, through the integration of sophisticated algorithms with legal acumen, this application illustrates how technology can enhance conventional legal methods.

\section{Related Work}\label{sec2}
Over the years due to rapid increase in legal cases and the introduction of new sections have caused trouble for legal practitioners in their research or retrieving citations for their work. This involved exhaustive manual work and was a tedious process. With many legal documents now being electronically stored various methods can be implemented to automate this process time efficiently \cite{chhatwal2018explainable}\cite{wei2018empirical}\cite{matthijssen1995intelligent}

Kumar, S., Reddy, P. K., Reddy, V. B., \& Singh \cite{kumar2011similarity} proposed four methods to evaluate the similarity between the legal judgements that involved the usage of keywords. The first method is All-term cosine similarity in which each text is treated as a vector of terms and the similarity is calculated based on TF-IDF scores. In Legal-term cosine similarity, a legal dictionary is included which helped improve the relevance. The legal-term cosine similarity method performed better than the all-term method, indicating that domain-specific terms are more effective. In bibliographic coupling and co-citation methods, the similarity is measured based on common references cited in the judgements and common citations received by the judgements respectively.

Biagioli, C., Francesconi, E., Passerini, A., Montemagni, S., \& Soria, C \cite{biagioli2005automatic} evaluates two key modules namely Automatic Provision Classifier and Provision Argument Extractor. The Automatic Provision Classifier was tested on a dataset consisting of paragraphs from Italian legislative texts. Improvement in the performance was observed after the dataset was preprocessed which mainly included the removal of quoted sentences from the documents. Various document representation and feature selection strategies were explored, such as term weighting schemes and feature selection methods. An accuracy of 92.44\% was achieved using a combination of preprocessing steps, stemming, and specific feature selection strategies. The Provision Argument extractor operates in two stages: syntactic pre-processing and semantic annotation. Syntactic pre-processing involves tokenization, normalization, morphological analysis, lemmatization, POS tagging, and shallow parsing. Semantic annotation identifies relevant arguments. An evaluation was conducted on a subset of the dataset used for the Automatic Classifier. The results achieved an overall success rate of 82.09\%.

\section{Judgement Citations Extraction}\label{sec3}
Legal practitioners often use citations in their work therefore automating this process without too much load on the users is necessary such that it is time efficient. We propose a methodology where whenever a user uploads a case description, we automatically extract citations that are contextually related to it.
\begin{figure}[H]
    \centering
    \includegraphics[width=1\linewidth]{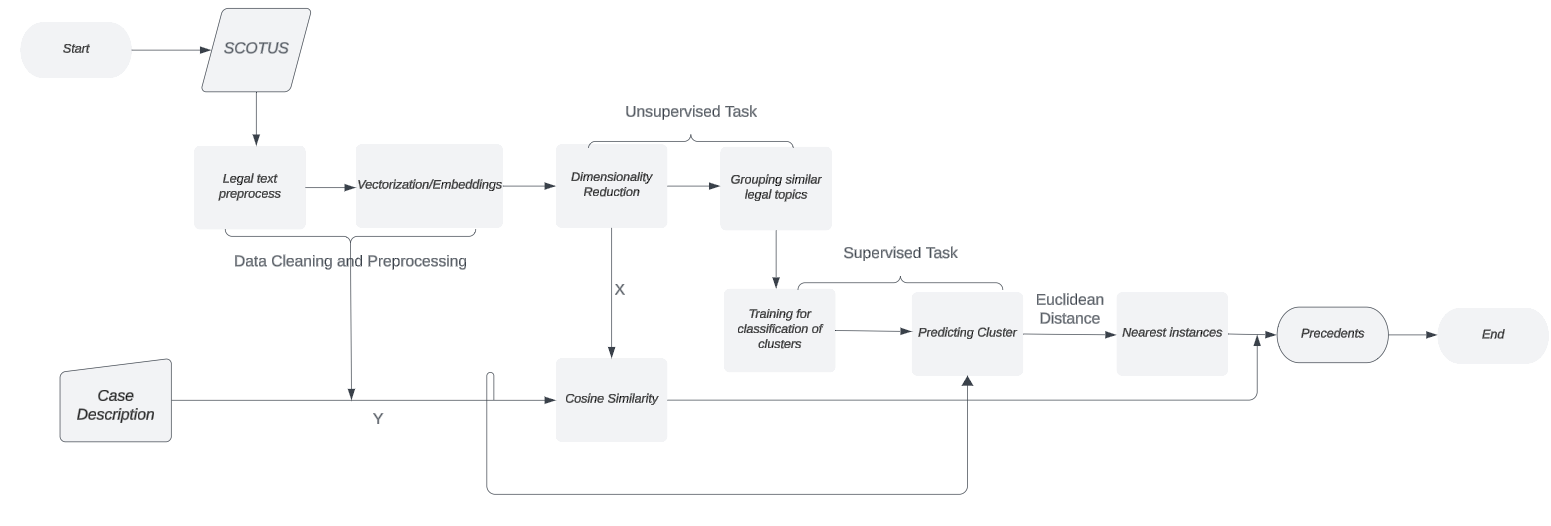}
    \caption{Working of Citation Retrieval Application}
    \label{fig:imagejc1}
\end{figure}
Particularly the application extracts 5 citations out of which the most similar citation is given by Cosine Similarity \cite{gunawan2018implementation} and the other 4 similar citations are given by classification algorithms. Since labels are initially not present, the classification task is not directly possible, therefore it is necessary to perform clustering and assign each case description to a cluster with a number as a label and use these to classify. In \cref{fig:imagejc1}, the whole judgement citation extraction process is outlined and in the following sections, we discuss it in detail.
\subsection{SCOTUS Dataset}\label{subsec1}

The data used in the proposed work is The Supreme Court of The United States (SCOTUS) dataset but the methodology proposed can work on any dataset. This dataset contains most of the information that a legal practitioner would require. Here, we understand the requirements and information that an attorney will need before searching for citations and therefore our perspective of the dataset depends on the necessities of an attorney.

\begin{figure}[H]
    \centering
    \includegraphics[width=1\linewidth]{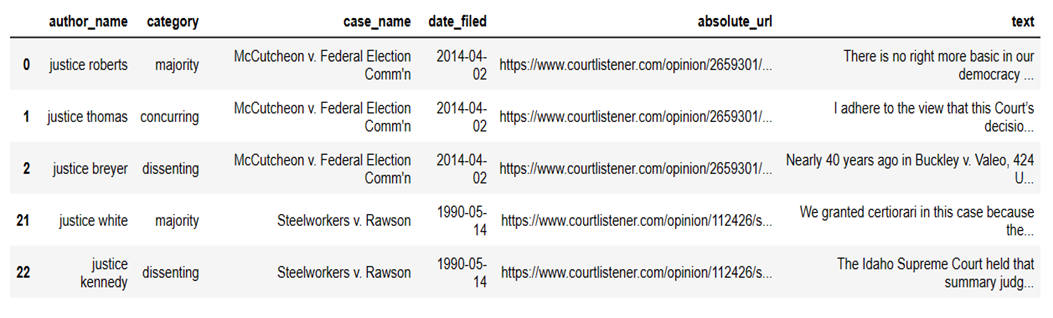}
    \caption{Overview of SCOTUS Dataset}
    \label{fig:imagejc2}
\end{figure}

The dataset consists of 14 features that include important information, the summary of the 7 main features, shown in \cref{fig:imagejc2} is as follows:
\begin{enumerate}[i]
    \item Justice name: Name of the Justice or author
    \item Case Name:  Tells the parties or people involved in the case
    \item Case Description: The full-length description of a case or opinion.
    \item The year in which the case was filed
    \item The URL or link to the CourtListener Database
    \item Case categories such as majority, dissenting, second dissenting, per-curiam
    \item The ID of the case in the Supreme Court Database
\end{enumerate}
And other features like case direction, votes, etc.
The cases in the data are from the year 1700s since there have been a lot of changes in the legal domain such as the introduction and scraping of sections. This might lead to new problems when capturing the semantics, therefore our methodology only consists of cases from the year 1985 to keep it the latest. The final dataset consists of 5880 records. Although a lot of information is provided through the features in the data, the primary focus is on the case description or the legal opinion that provides the entire description. We assume that it is the most important aspect of the citation retrieval.

\subsection{Preprocessing the Case Descriptions}\label{subsec2}
\begin{figure}[H]
    \centering
    \includegraphics[width=1\linewidth]{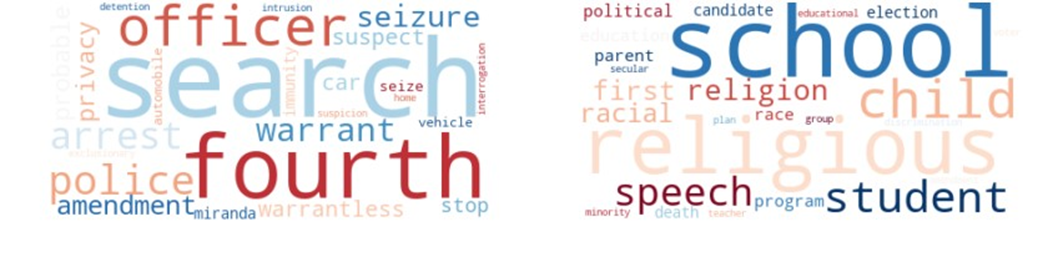}
    \caption{Word Clouds of Case Descriptions}
    \label{fig:imagejc3}
\end{figure}

\begin{figure}[H]
    \centering
    \includegraphics[width=1\linewidth]{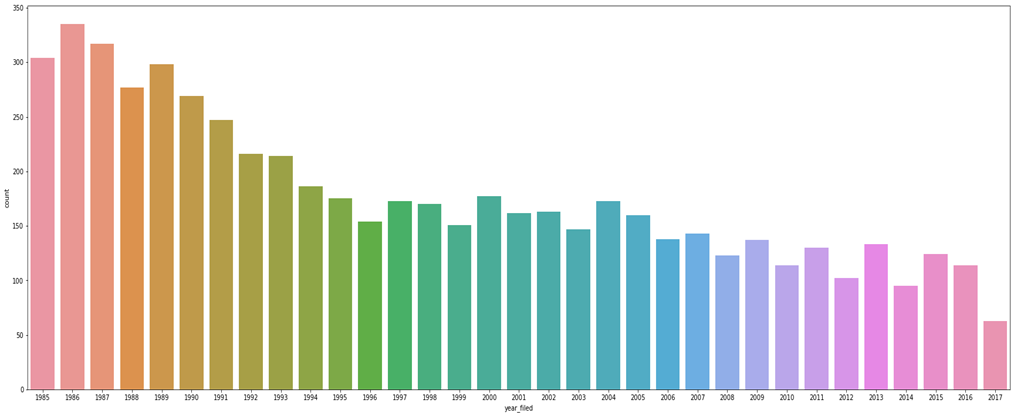}
    \caption{Number of Cases per year}
    \label{fig:imagejc4}
\end{figure}

\begin{figure}[H]
    \centering
    \includegraphics[width=0.75\linewidth]{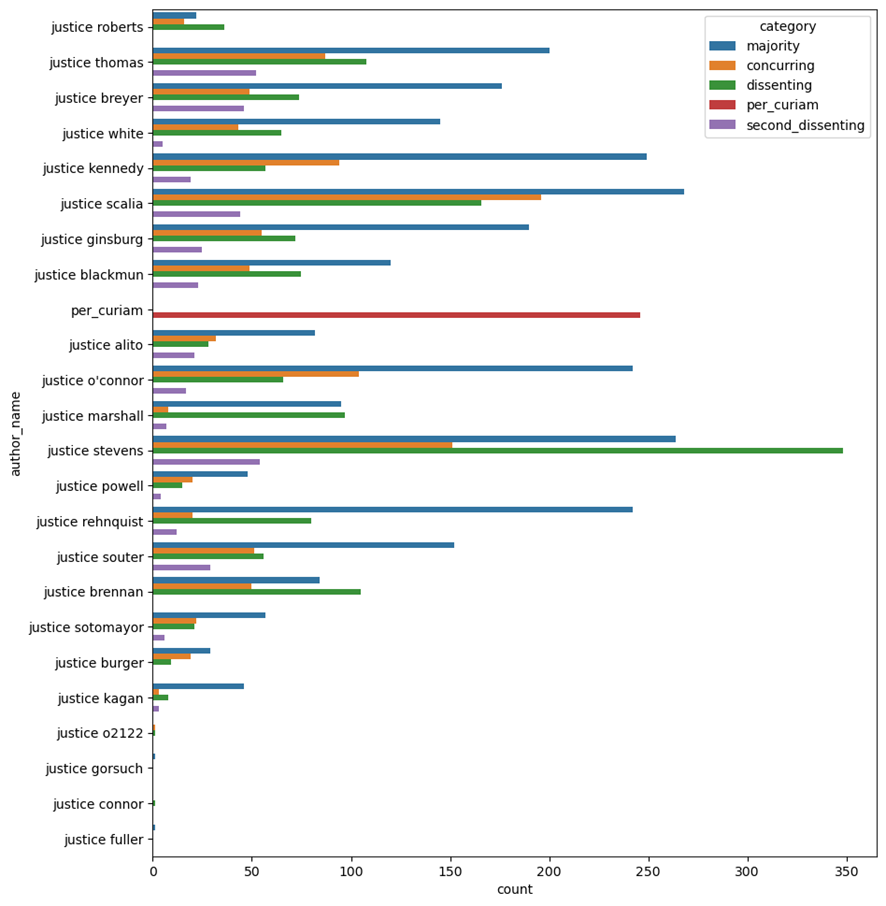}
    \caption{Number of Cases taken by each Judge}
    \label{fig:imagejc5}
\end{figure}

\begin{figure}[H]
    \centering
    \includegraphics[width=1\linewidth]{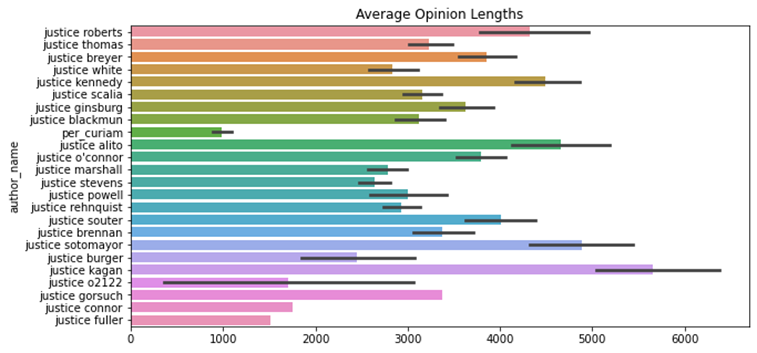}
    \caption{Average Opinion lengths of each Judge}
    \label{fig:imagejc6}
\end{figure}

The first step that needs to be done before diving into modeling is the preprocessing of the case description. These descriptions must be converted into numeric vectors, as machine learning algorithms only work with numbers.  The main concept here is to remove noise present in the text data, including stop words elimination, converting numbers to words as preserving section numbers is important, punctuation, contractions like question tags, and lemmatization, very short Case Descriptions were also removed as these descriptions didn’t provide any significance. Exploratory Data Analysis (EDA) done in this step here is shown in \cref{fig:imagejc3}, \cref{fig:imagejc4}, \cref{fig:imagejc5}
From the \cref{fig:imagejc4}, it is evident that the number of cases filed per year decreased from the year 1985 to the year 2017. Further, since we discarded very short case descriptions, the length of each case description given by different judges is necessary to analyze, this is shown in \cref{fig:imagejc6}

We implemented two approaches, the first one is Latent Semantic Analysis (LSA) \cite{landauer1998introduction} which follows the traditional Term Frequency - Inverse Document Frequency (TF-IDF) \cite{alobaydy2022document}, When it comes to finding citations the combination of LSA and TF IDF is extremely valuable. TF IDF helps identify keywords and phrases in documents by assigning weights based on their significance. LSA goes further by exploring the meaning and relationships, between terms and documents. This approach allows for the retrieval of citations that go beyond keyword matches as it finds cases with shared contexts. LSA's ability to reduce complexity while maintaining content improves the accuracy and relevance of finding citations in the field.

The second approach of text preprocessing is Sentence embedding, sentence embeddings play a role in improving the retrieval of citations, in the legal domain. Universal Sentence Encoder (USE) \cite{cer2018universal} which is the state-of-art sentence embedding model is used to extract contextual embeddings for all the case descriptions given. USE is a deep learning model that has been pre-trained to excel at converting text, here case descriptions, into vector representations that capture their semantic meaning. The USE produces vectors of 512 dimensions for each input sentence or text ensuring vector lengths that facilitate effortless and effective comparisons of similarity as well as text classification tasks.

\subsection{Clustering the Case Descriptions (The Unsupervised Task)}\label{subsec3}

After converting the legal case descriptions as vectors from the above-mentioned two approaches, the next step is to cluster or group these case descriptions \cite{lu2011legal}. The idea behind this is, that since the vectors capture the semantics of the given case description, they are grouped based on their semantic similarity. The concept says that when two vectors are closer to each other, they have some semantic association, meaning that the vectors that have similar descriptions are grouped together and therefore are pertinent to each other. This can be done using unsupervised clustering tasks for machine learning algorithms, we experimented with this using the clustering algorithm K-Means Clustering \cite{duo2021kmeans}\cite{alobaydy2022document} and Density-based Spatial Clustering of Applications with Noise (DBSCAN) \cite{deng2020dbscan}, and the number of clusters was determined using evaluation metrics that tell us the optimal number of clusters in these two algorithms Elbow Method and Silhouette Score.

The Elbow Method \cite{cui2020introduction} is used to determine the optimal number of clusters in a dataset by plotting the sum of squared distances from each point to its assigned cluster center (within-cluster sum of squares, WCSS) against the number of clusters. As the number of clusters increases, the WCSS decreases, but the rate of decrease slows down; the "elbow" point on the plot indicates a balance between the number of clusters and the variance explained by the clusters. The optimal number of clusters is typically at the "elbow" point where the WCSS starts to diminish slower, suggesting that additional clusters do not provide significant improvements. Silhouette scores \cite{shahapure2020cluster} measure the quality of clustering by evaluating how similar an object is to its cluster (cohesion) compared to other clusters (separation), producing a score between -1 and 1. A high silhouette score indicates that objects are well-matched to their cluster and poorly matched to neighboring clusters, suggesting a well-defined cluster structure. Average silhouette scores for different numbers of clusters can be plotted, with the optimal number of clusters corresponding to the highest average silhouette score, indicating the best separation and cohesion, where an object or a point belongs to one case description in the current work

For K-Means, The K value was tested from 2 to up to 100 but there was no significant difference from 50 to 100 and are therefore omitted. Due to a lack of legal expert advice, the clusters formed are ambiguous and so the metrics do not accurately mark the K value, such advice can be used to rank and mark the falsely clustered descriptions therefore making them more robust. This was tested on both the TF-IDF + LSA method and the Sentence Embedding method, the number of clusters formed was 47 in the former case and 26 in the latter case. 

In LSA, the dimensions are reduced from 1,04,272 features acquired from TF-IDF to 4500 features, i.e. length of each vector that captured 97.8\% of variability in the data.
\begin{figure}[htbp]
    \centering
    \begin{minipage}[b]{0.45\textwidth}
        \centering
        \includegraphics[width=0.75\linewidth]{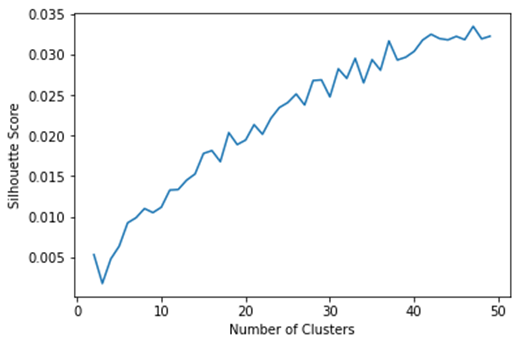}
        \caption{Silhouette Scores for TF-IDF + LSA approach}
        \label{fig:imagejc43b}
    \end{minipage}
    \hspace{0.05\textwidth} 
    \begin{minipage}[b]{0.45\textwidth}
        \centering
        \includegraphics[width=0.75\linewidth]{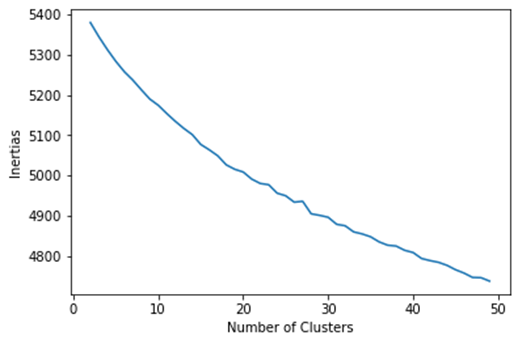}
        \caption{Elbow Graph for TF-IDF + LSA approach}
        \label{fig:imagejc43c}
    \end{minipage}
\end{figure}
Since K-Means is our primary approach in this step and DBSCAN produced results not much different than K-Means, the following \cref{fig:imagejc43b}, \cref{fig:imagejc43c} show the Silhouette score and Elbow method graphs for TF-IDF + LSA approach respectively.

\begin{figure}[htbp]
    \centering
    \begin{minipage}[b]{0.45\textwidth}
        \centering
        \includegraphics[width=0.75\linewidth]{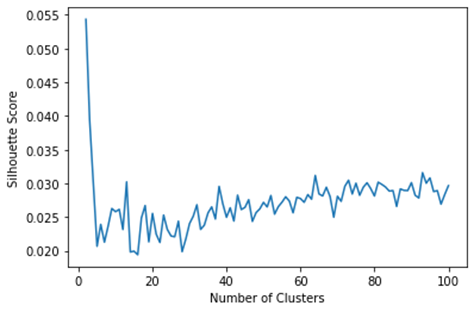}
        \caption{Silhouette Scores for Sentence embedding approach}
        \label{fig:imagejc4d}
    \end{minipage}
    \hspace{0.05\textwidth} 
    \begin{minipage}[b]{0.45\textwidth}
        \centering
        \includegraphics[width=0.75\linewidth]{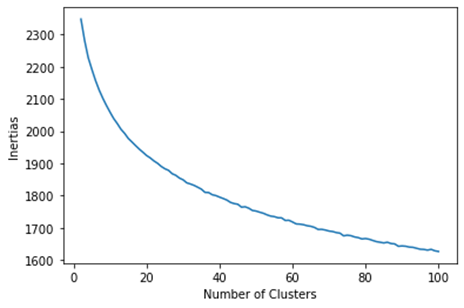}
        \caption{Elbow Graph for Sentence embedding approach}
        \label{fig:imagejc4e}
    \end{minipage}
\end{figure}

In the second approach which is Sentence Embeddings, the output vector is a standard 512 vector that has the relevant semantic information given by USE, on clustering these embeddings, the evaluation metrics are given in \cref{fig:imagejc4d}, \cref{fig:imagejc4e}
Even though the clusters formed are unstable, the number of clusters formed for 2 approaches are given in Table \ref{tab:clusters}. The first one, K-Means and DBSCAN with TF-IDF + LSA is 47 clusters, and the second one, K-Means and DBSCAN with embeddings are 26 clusters.

\begin{figure}[htbp]
    \centering
    \begin{minipage}[b]{0.45\textwidth}
        \centering
        \includegraphics[width=1\linewidth]{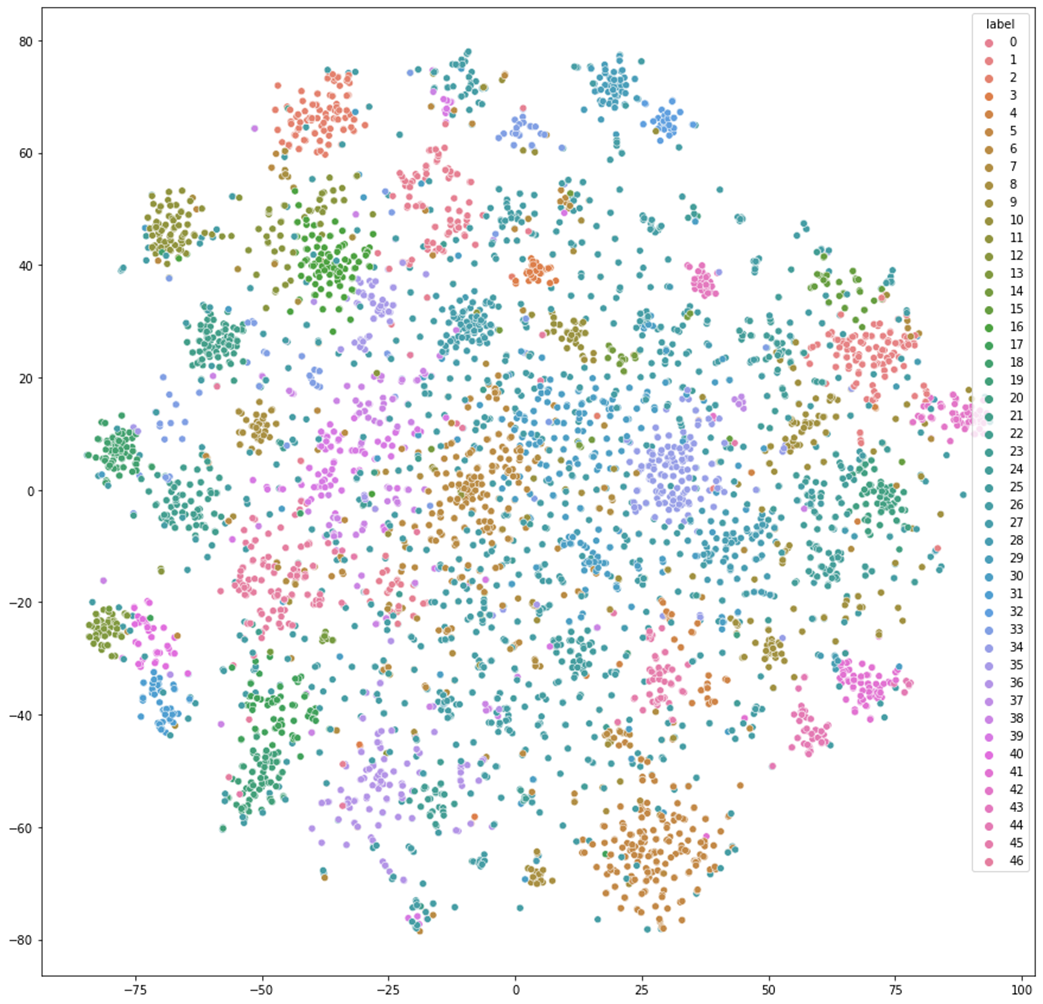}
        \caption{Data after assigning clusters for TF-IDF + LSA}
        \label{fig:imagejc43f}
    \end{minipage}
    \hspace{0.05\textwidth} 
    \begin{minipage}[b]{0.45\textwidth}
        \centering
        \includegraphics[width=1\linewidth]{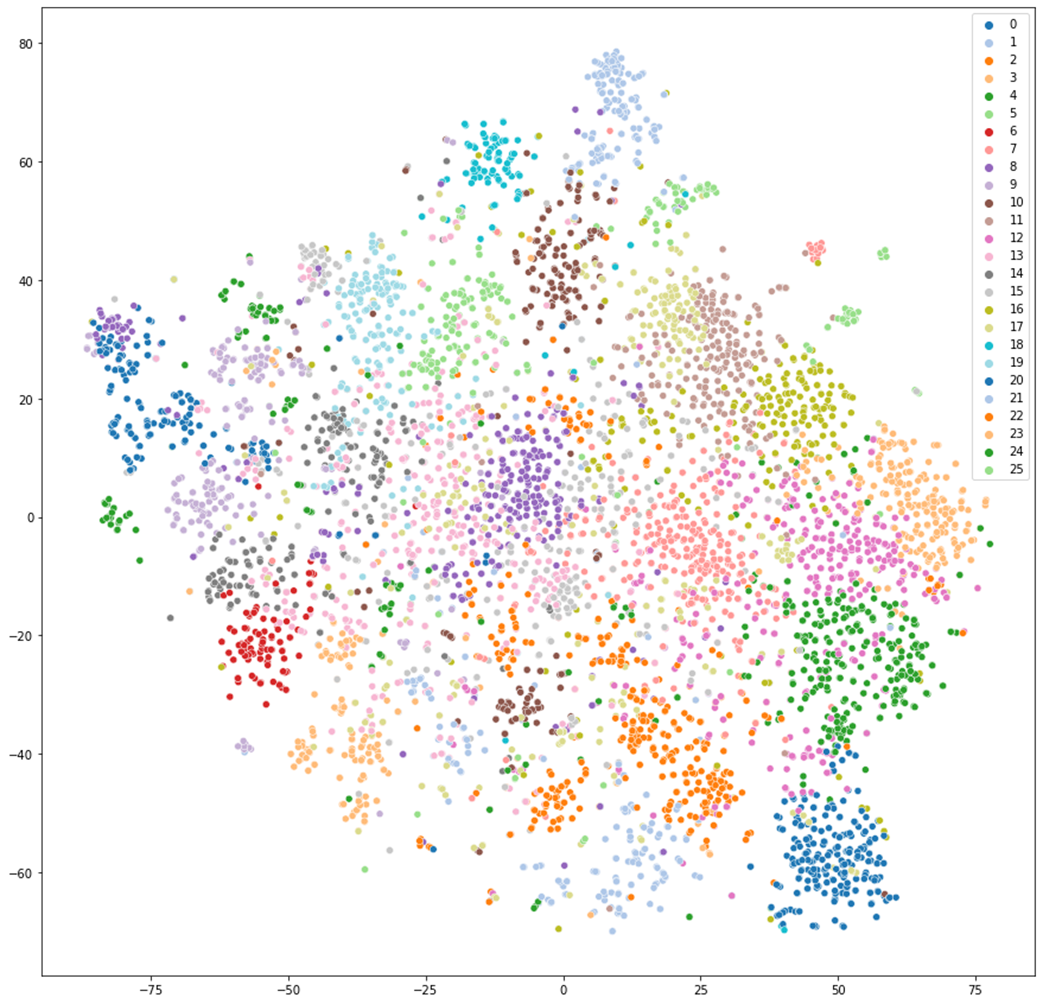}
        \caption{Data after assigning clusters for sentence embeddings}
        \label{fig:imagejc43g}
    \end{minipage}
\end{figure}

\begin{table}[h!]
\centering
\caption{Clusters formed by two approaches}
\label{tab:clusters}
\begin{tabular}{|c|c|c|}
\hline
 & \textbf{TF-IDF + LSA} & \textbf{Embedding} \\ 
\hline
\textbf{KMeans} & 47 & 26 \\ 
\hline
\textbf{DBSCAN} & 47 & -- \\ 
\hline
\end{tabular}
\end{table}
Finally, in \cref{fig:imagejc43f} and \cref{fig:imagejc43g} shows how the data looks after assigning the clusters to each case description in the corpus two-dimensionally, done using t-distributed Stochastic Neighbor Embedding (t-SNE) \cite{van2008visualizing}.
Labels are added to each case description in the dataset, here label is the cluster number to which the respective case description belongs to. Therefore, the output of this application is now the input for the next step.
\subsection{Retrieving the Precedents (The Supervised Task)}\label{subsec4}

From the previous step, we have case descriptions (features) and the clusters they belong to (labels). With these labels, supervised task \cite{stow2023improved} can be performed by dividing them into train and test sets and feeding them into the Machine Learning model. This step aims to train a model such that it can predict the label of the new case description given by a legal practitioner which will facilitate the retrieval of citations to that particular case. The training and testing sets were divided into 67\% and 33\% respectively.

In this step, three supervised models were implemented. The first one is K Nearest Neighbors (KNN), KNN works on distance-based approaches and since similarity is our focus from the beginning, KNN is suitable for the task at hand, it takes K opinions in the dataset that is nearest to the new description given and determines the label by taking the majority of the labels that near to it. We experimented with this with K values ranging from 3 to 49.
The second model used in this was an Artificial Neural Network, i.e. a simple Feed Forward Neural Network. The architecture is a sequential model of 3 Dense layers, an input layer with 4500 neurons in the case of TF-IDF + LSA and 512 in the case of sentence embeddings, a hidden layer with 128 neurons for both, the activation function used in input and hidden layer was ReLu and finally an output layer with the respective number of classes for each approach, 47 for TF-IDF + LSA and 26 for Sentence Embeddings.

The final model used in the current work was Support Vector Machines (SVM). SVM is best for higher dimensional data and in the current work, we have at least 512 dimensions to work on. It defines hyperplanes that separate the classes. Grid search was employed to identify the right parameters such as best kernel, number of misclassifications allowed, and a gamma value that is required to achieve the task. 

There is an additional method used here, irrespective of the labels assigned, as there are some apparent issues that are seen in the metrics of the number of clusters formed which is the number of clusters not being stable and the cluster metrics not giving us a valid number of clusters, Cosine Similarity approach was also employed to overcome the issues in the clustering step and possible issues that might occur in classification task for example, since the legal domain is sensitive, suppose the highest accuracy of the classification task is 90\%, cosine similarity answers this 10\% of uncertainty. For a new case description, it is converted to a vector, and with all the legal descriptions as vectors in our database, the most relevant case is retrieved using the Cosine Similarity.
The other relevant cases are retrieved using Clustering and Classification tasks. After predicting the cluster to which the new description belongs, Euclidean distance is used to retrieve legal opinions that are nearest to it in that cluster and these opinions will work as citations to the new legal case

\section{Results and Discussions}\label{sec4}
\begin{figure}[H]
    \centering
    \includegraphics[width=1\linewidth]{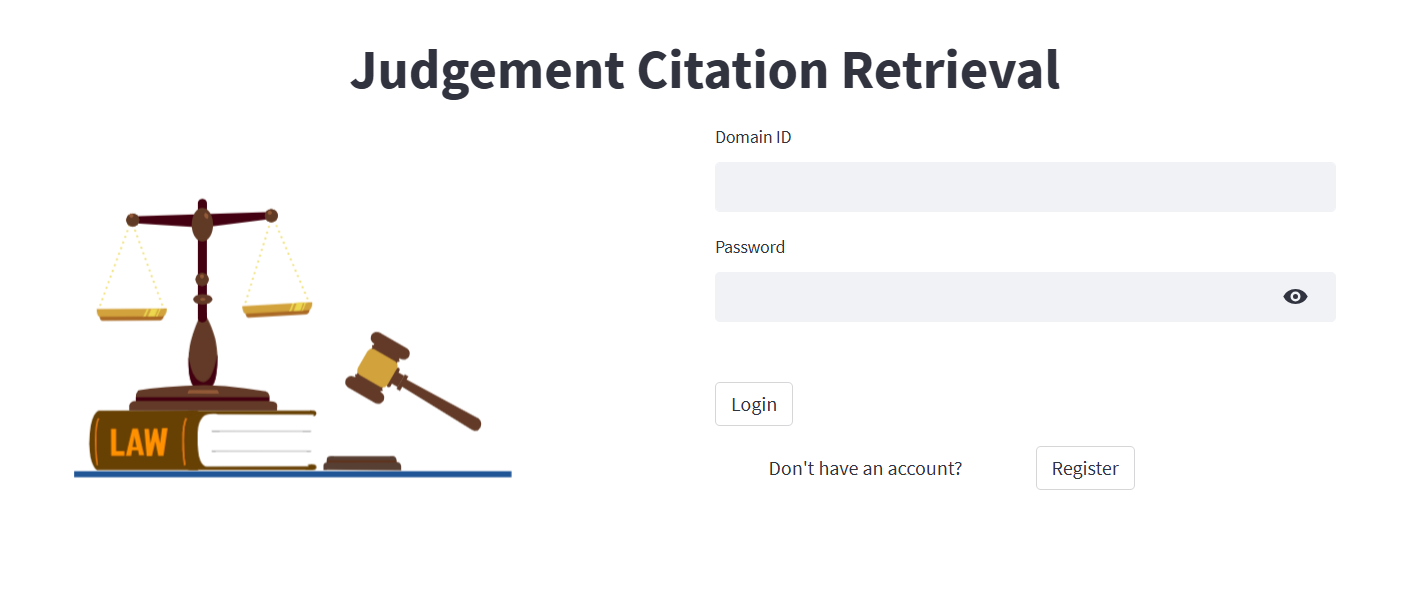}
    \caption{Application Login Page}
    \label{fig:imagejc5a}
\end{figure}

\begin{figure}[H]
    \centering
    \includegraphics[width=1\linewidth]{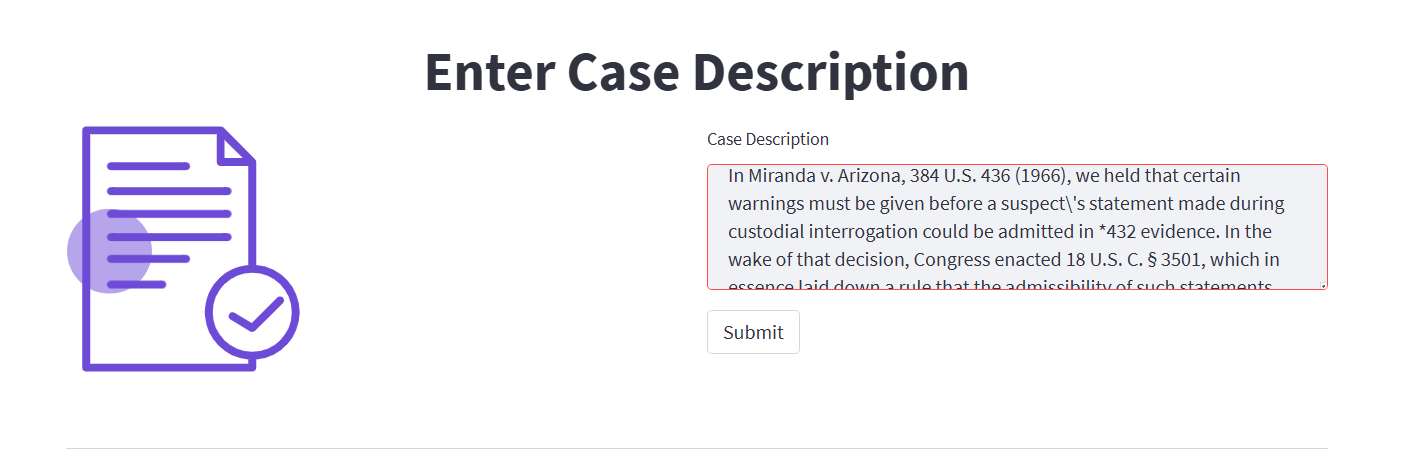}
    \caption{Enter Case Description}
    \label{fig:imagejc5b}
\end{figure}

\begin{figure}[H]
    \centering
    \includegraphics[width=1\linewidth]{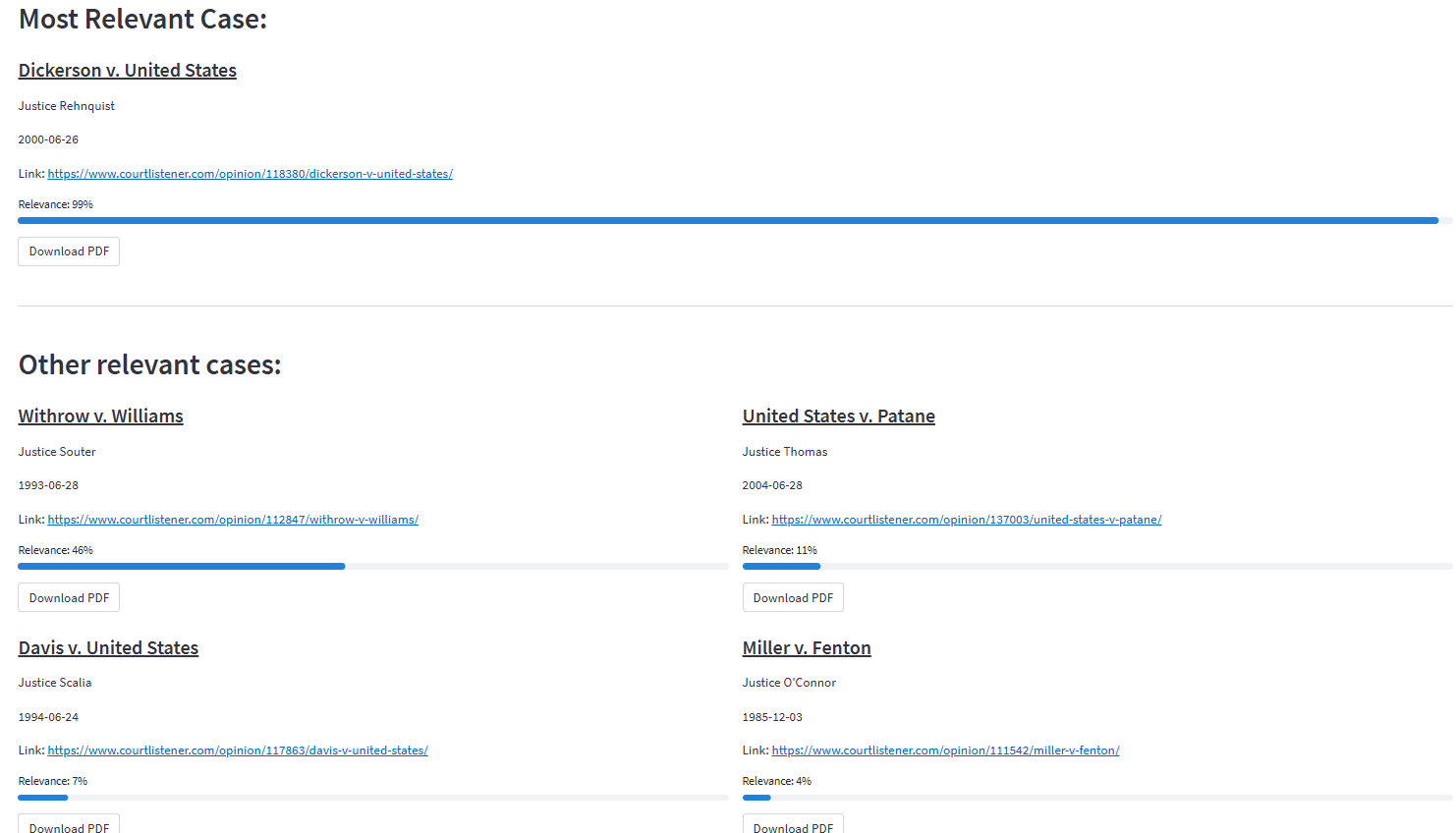}
    \caption{Retrieved Cases}
    \label{fig:imagejc5c}
\end{figure}

\begin{figure}[H]
    \centering
    \includegraphics[width=0.75\linewidth]{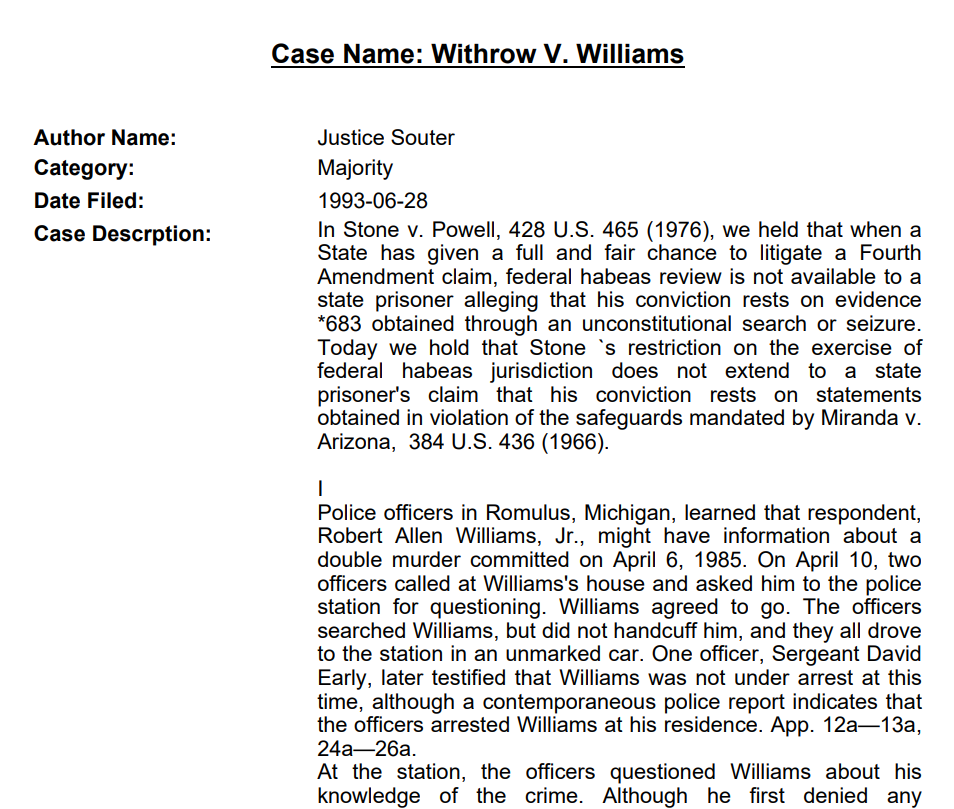}
    \caption{Downloaded PDF}
    \label{fig:imagejc5d}
\end{figure}

Legal Case Descriptions are classified into different classes where the class names are numbers assigned to the clusters, both approaches, sentence embeddings, and LSA were tested in the classification step and the highest accuracy obtained was 90.9\% from  Artificial Neural Networks (ANN) combined with KMeans + Embedding. The accuracy obtained for each approach with each classification method is shown in Table \ref{tab:results}. The least accuracies obtained were of methods that involved K Nearest Neighbors (KNN). The following table shows the accuracies obtained in each method. It was observed that although a pre-trained model USE had to be loaded from Tensorflow Hub, the methods that used Embeddings worked faster compared to the LSA approach. The former retrieved the cases within seconds but the latter took approximately 1 minute to retrieve which is subject to change based on the length of the case description given. The lengths given case descriptions varied from 4 pages when converted to a PDF to more than 30 pages. 
\begin{table}[h!]
\centering
\caption{Performance comparison of different classifiers}
\label{tab:results}
\begin{tabular}{|c|c|c|}
\hline
 & \textbf{KMeans + TF-IDF + LSA} & \textbf{KMeans + Embedding} \\ 
\hline
\textbf{SVM} & 85.01\% & 89.8\% \\ 
\hline
\textbf{ANN} & 84.14\% & 90.9\% \\ 
\hline
\textbf{KNN} & $\leq$74\% & $\leq$75\% \\ 
\hline
\end{tabular}
\end{table}
The application can be accessed only by legal practitioners due to its domain sensitivity, therefore a user can access the application by logging in with their domain ids as shown in \cref{fig:imagejc5a}, although if the user is not present in the database they can register with their domain id.
Whenever a user prompts the application with a case description as shown in \cref{fig:imagejc5b} and clicks on submit
On clicking submit, the application starts the pipeline mentioned in \cref{fig:imagejc1} and extracts the most relevant and other relevant cases from the database as shown in \cref{fig:imagejc5c}, here there is a button called "Download PDF" and relevance bar. The relevance bar shows how similar the extracted case is to the given case.
The "Download PDF" button downloads the citation as a pdf with the necessary information in it as shown in \cref{fig:imagejc5d}
\section{Conclusion and Future Work}\label{sec5}
In conclusion, the current work marks a groundbreaking advancement in legal research by automating the text retrieval system using NLP that efficiently retrieves pertinent legal cases. It identifies cases time-efficiently, offering legal professionals a user-friendly interface that enhances the decision-making process by providing data-driven insights and evidence-based strategies. In pioneering legal analytics, this application has the potential to revolutionize legal practice, making it more informed, and empowering legal professionals with a transformative tool for their research and decision-making needs. 

Although this application promises good results, there are some limitations to it. Firstly, the proposed methodology works on generalized texts and not jargon related specifically to the legal domain. Having an embedding that is not generalized to other domains would understand and produce embeddings better than USE. Secondly, The clusters formed at the unsupervised step sometimes produce different numbers of clusters in different iterations this is due to uncertainty in some cases where a particular case might belong to more than one category such as abortion can be a right to privacy issue or a medical issue. There are many such cases where in which such uncertainty is found. To avoid these a legal expert’s advice might be crucial for this to work well. Lastly, the current methodology doesn’t assign a topic name to the cluster, NLP concepts like Topic Modeling \cite{carlsson2021classification} can be implemented on these clusters which enables a legal practitioner to understand to which topic the cluster belongs such as abuse, elections, etc. that case description might belong to but again this might also require a legal advice to work reliably for reasons discussed previously.

For legal researchers, to understand the legal jargon well Named Entity Recognition (NER) \cite{leitner2019fine} can be implemented which recognizes the in-depth jargon related to the case and can assist them in understanding this jargon. The Topic Modeling and NER can be useful when a pipeline is developed that generates legal documents related to a specific domain of the case, as legal documents generally vary with the type of case. Since many new sections are introduced and a lot of different legal cases are introduced these changes lead to data drift. This drift has to be identified and the model must be retrained frequently. For example, there are cases related to same-sex marriage in recent times which was not that common previously.


\bibliography{sn-bibliography}

\begin{thebibliography}{10}
\expandafter\ifx\csname url\endcsname\relax
  \def\url#1{\burl{#1}}\fi
\expandafter\ifx\csname urlprefix\endcsname\relax\def\urlprefix{URL }\fi
\providecommand{\bibinfo}[2]{#2}
\providecommand{\eprint}[2][]{\url{#2}}
\providecommand{\doi}[1]{\url{https://doi.org/#1}}
\bibcommenthead

\bibitem{moens1997abstracting}
\bibinfo{author}{Moens, M.~F.}, \bibinfo{author}{Uyttendaele, C.} \& \bibinfo{author}{Dumortier, J.}
\newblock \bibinfo{title}{Abstracting of legal cases: The salomon experience} (\bibinfo{year}{1997}).

\bibitem{raghav2015text}
\bibinfo{author}{Raghav, K.}, \bibinfo{author}{Balakrishna~Reddy, P.}, \bibinfo{author}{Balakista~Reddy, V.} \& \bibinfo{author}{Krishna~Reddy, P.}
\newblock \bibinfo{title}{Text and citations based cluster analysis of legal judgments} (\bibinfo{year}{2015}).

\bibitem{chhatwal2018explainable}
\bibinfo{author}{Chhatwal, R.} \emph{et~al.}
\newblock \bibinfo{title}{Explainable text classification in legal document review a case study of explainable predictive coding} (\bibinfo{year}{2018}).

\bibitem{wei2018empirical}
\bibinfo{author}{Wei, F.}, \bibinfo{author}{Qin, H.}, \bibinfo{author}{Ye, S.} \& \bibinfo{author}{Zhao, H.}
\newblock \bibinfo{title}{Empirical study of deep learning for text classification in legal document review} (\bibinfo{year}{2018}).

\bibitem{matthijssen1995intelligent}
\bibinfo{author}{Matthijssen, L.~J.}
\newblock \bibinfo{title}{An intelligent interface for legal databases} (\bibinfo{year}{1995}).

\bibitem{kumar2011similarity}
\bibinfo{author}{Kumar, S.}, \bibinfo{author}{Reddy, P.~K.}, \bibinfo{author}{Reddy, V.~B.} \& \bibinfo{author}{Singh, A.}
\newblock \bibinfo{title}{Similarity analysis of legal judgments} (\bibinfo{year}{2011}).

\bibitem{biagioli2005automatic}
\bibinfo{author}{Biagioli, C.}, \bibinfo{author}{Francesconi, E.}, \bibinfo{author}{Passerini, A.}, \bibinfo{author}{Montemagni, S.} \& \bibinfo{author}{Soria, C.}
\newblock \bibinfo{title}{Automatic semantics extraction in law documents} (\bibinfo{year}{2005}).

\bibitem{gunawan2018implementation}
\bibinfo{author}{Gunawan, D.}, \bibinfo{author}{Sembiring, C.~A.} \& \bibinfo{author}{Budiman, M.~A.}
\newblock \bibinfo{title}{The implementation of cosine similarity to calculate text relevance between two documents} (\bibinfo{year}{2018}).

\bibitem{landauer1998introduction}
\bibinfo{author}{Landauer, T.~K.}, \bibinfo{author}{Foltz, P.~W.} \& \bibinfo{author}{Laham, D.}
\newblock \bibinfo{title}{An introduction to latent semantic analysis} (\bibinfo{year}{1998}).

\bibitem{alobaydy2022document}
\bibinfo{author}{Al-Obaydy, W.~I.}, \bibinfo{author}{Hashim, H.~A.}, \bibinfo{author}{Najm, Y.~A.} \& \bibinfo{author}{Jalal, A.~A.}
\newblock \bibinfo{title}{Document classification using term frequency-inverse document frequency and k-means clustering} (\bibinfo{year}{2022}).

\bibitem{cer2018universal}
\bibinfo{author}{Cer, D.} \emph{et~al.}
\newblock \bibinfo{title}{Universal sentence encoder} (\bibinfo{year}{2018}).

\bibitem{lu2011legal}
\bibinfo{author}{Lu, Q.}, \bibinfo{author}{Conrad, J.~G.}, \bibinfo{author}{Al-Kofahi, K.} \& \bibinfo{author}{Keenan, W.}
\newblock \bibinfo{title}{Legal document clustering with built-in topic segmentation} (\bibinfo{year}{2011}).

\bibitem{duo2021kmeans}
\bibinfo{author}{Duo, J.}, \bibinfo{author}{Zhang, P.} \& \bibinfo{author}{Hao, L.}
\newblock \bibinfo{title}{A k-means text clustering algorithm based on subject feature vector} (\bibinfo{year}{2021}).

\bibitem{deng2020dbscan}
\bibinfo{author}{Deng, D.}
\newblock \bibinfo{title}{Dbscan clustering algorithm based on density} (\bibinfo{year}{2020}).

\bibitem{cui2020introduction}
\bibinfo{author}{Cui, M.}
\newblock \bibinfo{title}{Introduction to the k-means clustering algorithm based on the elbow method} (\bibinfo{year}{2020}).

\bibitem{shahapure2020cluster}
\bibinfo{author}{Shahapure, K.~R.} \& \bibinfo{author}{Nicholas, C.}
\newblock \bibinfo{title}{Cluster quality analysis using silhouette score} (\bibinfo{year}{2020}).

\bibitem{van2008visualizing}
\bibinfo{author}{Van~der Maaten, L.} \& \bibinfo{author}{Hinton, G.}
\newblock \bibinfo{title}{Visualizing data using t-sne} (\bibinfo{year}{2008}).

\bibitem{stow2023improved}
\bibinfo{author}{Stow, M.~T.}, \bibinfo{author}{Ugwu, C.} \& \bibinfo{author}{Onyejegbu, L.}
\newblock \bibinfo{title}{An improved model for legal case text document classification} (\bibinfo{year}{2023}).

\bibitem{carlsson2021classification}
\bibinfo{author}{Carlsson, H.} \& \bibinfo{author}{Lindgren, T.}
\newblock \bibinfo{title}{Classification of legal documents a topic modeling approach} (\bibinfo{year}{2021}).

\bibitem{leitner2019fine}
\bibinfo{author}{Leitner, E.}, \bibinfo{author}{Rehm, G.} \& \bibinfo{author}{Moreno-Schneider, J.}
\newblock \bibinfo{title}{Fine-grained named entity recognition in legal documents} (\bibinfo{year}{2019}).

\end{thebibliography}

\end{document}